\documentclass{bioinfo}
\copyrightyear{2012}
\pubyear{2012}

\begin{document}
\firstpage{1}

\title[SSW Library]{SSW Library: An SIMD Smith-Waterman C/C++ Library for Use in Genomic Applications}
\author[M.Zhao \textit{et~al}]{Mengyao Zhao\,, Wan-Ping Lee\,, Erik Garrison\, and Gabor T. Marth\footnote{To whom correspondence should be addressed.}\, }
\address{Department of Biology, Boston College, 140 Commonwealth Avenue, Chestnut Hill, MA 02467, USA}

\history{Received on XXXXX; revised on XXXXX; accepted on XXXXX}

\editor{Associate Editor: XXXXXXX}

\maketitle

\begin{abstract}

\section{Summary:}
The Smith-Waterman (SW) algorithm, which produces the optimal pairwise alignment between two sequences, is frequently used as a key component of fast heuristic read mapping and variation detection tools, but current implementations are either designed as monolithic protein database searching tools or are embedded into other tools. To facilitate easy integration of the fast Single-Instruction-Multiple-Data (SIMD) SW algorithm into third-party software, we wrote a C/C++ library, which extends Farrar's Striped SW (SSW) to return alignment information in addition to the optimal SW score.
\section{Availability:}
SSW is available both as a C/C++ software library, as well as a stand-alone alignment tool wrapping the library's functionality at: 
https://github.com/mengyao/Complete-Striped- Smith-Waterman-Library
\section{Contact:} marth@bc.edu
\section{Supplementary Information:} 
\end{abstract}

\section{Introduction}
The Smith-Waterman-Gotoh algorithm (SW) \citep{Smith:1981fk, Gotoh:1982fk} is the most influential algorithm for aligning a pair of sequences. It is an essential component of the majority of aligners from the classical BLAST \citep{blast} to the more recent Bowtie2 \citep{Langmead:2012kx}. Although most of these aligners do not use SW directly to align a sequence to the whole genome sequence due to the quadratic time complexity of SW, they extensively use it for seed extension and for constructing the final alignment, and spend significant amount of CPU time on this algorithm. Due to the critical role of SW, many efforts have been made to accelerate SW, taking the advantages of special hardware such as single instruction multiple data (SIMD), field-programmable gate array (FPGA) and graphics processing unit (GPU). Among the three, SIMD based algorithms are most frequently used because they are compatible with most modern x86 CPUs. SIMD acceleration methods can be further divided into intra-sequence parallelization \citep{Wozniak:1997fk} and inter-sequence parallelization \citep{Rognes:2011fk}. Inter-sequence parallelization is only useful when many pairs of sequencing reads are aligned simultaneously; intra-sequence parallelization parallelizes for each single pairwise alignment, so it can be used more flexibly in various applications such as that needs aligning a single read against a potentially large genome reference sequence. Farrar's Striped SW \citep{Farrar:2007uq} with SWPS3's \citep{Szalkowski:2008fk} improvement is the fastest intra-sequence parallelized SIMD implementation running on x86 processors with the SSE2 instruction set. Indeed, Farrar's algorithm has been embedded in several popular genomic sequence mapping tools, such as BWA-SW \citep{Li:2010uq}, Bowtie2 \citep{Langmead:2012kx}, Novoalign (http://www.novocraft.com/) and Stampy \citep{Lunter:2011vn}.\\
\indent On the other hand, although striped SW is tens of times faster than a standard SW implementation, only a few aligners have used this more advanced algorithm. There are several practical obstacles. Firstly, implementing striped SW requires good understanding of SSE2 instructions and the more complex algorithm, which may take significant development time. Secondly, the original striped SW only gives the optimal alignment score but does not report the position or the detailed alignment, the information necessary for using SW as a component to construct the final alignment. How to report the position and alignment without affecting speed is non-trivial. Thirdly, while a few implementations report position and alignment, they are tightly integrated in a larger project and cannot be easily used in other programs. At last, when aligning a short read against a long sequence, we would like to know suboptimal alignments such that we can tell if the optimal position is trustworthy. Most existing libraries have not addressed this issue.\\
\indent Although striped SW has been published for five years, we are still in lack of a fast, versatile and standalone library, which leads us to the development of SSW Library, a small but comprehensive C/C++ library for pairwise sequence alignment with the striped SW algorithm.
\begin{methods}
\section{Algorithm and implementation}
Our algorithmic improvements focused on speeding up the Farrar's implementation and gaining access to the optimal alignment (in addition to the optimal achievable score), as well as the score of the best secondary alignment. For speedup, we adopted the ``lazy F loop'' improvement proposed by SWPS3 \citep{Szalkowski:2008fk}. To obtain additional alignment information while doesn't slow down the original algorithm: (1) we record the optimal alignment ending positions during the SIMD SW calculation and generate the detailed alignment by a reversed SIMD SW and a banded SW. When the score matrix is filled by the SIMD SW calculation, we store the maximal score of each column in a ``max'' array and record the complete column that has the maximal score of the whole matrix. Next, we locate the optimal alignment ending position on the reference and the query by seeking the maximal score in the array and the recorded column respectively. The reversed SIMD SW locates the best alignment beginning position from the ending position by calculating a much smaller scoring matrix. Then, the banded SW (whose band is defined by the beginning and ending positions) generates the detailed alignment. Since the alignment generation using the reversed SIMD SW and the banded SW only calculate a very small portion of the whole SW scoring matrix, we minimized the corresponding time cost. (2) We determine the secondary alignment score by seeking the second largest score in the ``max'' array. To avoid a proper sub-alignment of the primary alignment returned, we mask the elements in the region of the primary alignment of the ``max'' array and locate the second largest score from the unmasked elements (Figure \ref{fig1}). As a crucial step for locating the alignment position and estimating the suboptimal score, the ``max'' array generation is completed by adding an SSE2 command in the inner loop of Farrar's implementation and another in the outer loop, so that this additional time consumption is also limited.
\begin{figure}[!tpb]
\includegraphics[width=0.48\textwidth]{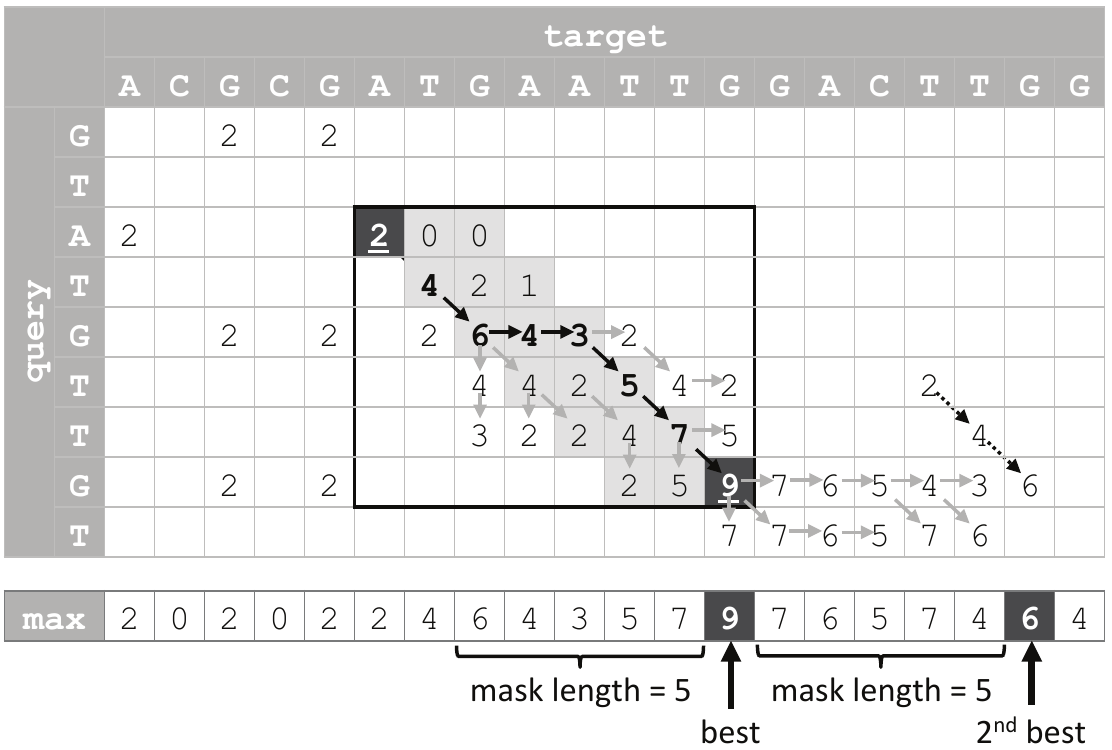}
\caption{Illustration of alignment traceback and suboptimal alignment score determination. An example SW score matrix is shown (penalties for match, mismatch, gap open and extension are 2, -1, -2, and -1 resp.). The bottom row indicates the maximum score for each column. The algorithm locates the optimal alignment ending position (the black cell with score 9) using the array of maximum scores, and then traces back to the alignment start position (the black cell with score 2) by searching a much smaller, locally computed score matrix (circled by the black rectangle). Finally, a banded SW calculates the detailed alignment by searching the shaded sub-region. The scores connected by solid arrows belong to the optimal alignment. The max array records the largest score of each column. After the optimal alignment score (marked by ``best'') is found, its neighborhood is masked, and the second largest score is reported outside the masked region (marked by 2nd best). The scores connected by dashed-line arrows trace the suboptimal alignment.}\label{fig1}
\end{figure}
\end{methods}
\section{Results}
\subsection{Usage}
The SSW library is an application program interface (API) that can be used as a component of C/C++ software to perform optimal protein or genome sequence alignment. The library returns the SW score, alignment location and traceback of the optimal alignment, and the alignment score and location of the suboptimal alignment. We provide the library with an executable alignment tool that can be used directly to perform protein or DNA alignments. It is a demonstration of the API usage and a practical tool for accurate whole viral or bacterial genome alignment. Moreover, since this tool is sufficiently fast and memory-efficient for alignment to very large reference genome sequences, e.g. the human genome, it can also be used to validate alignments produced by heuristic read mappers.
\subsection{Performance}
\begin{figure}[!tpb]
\includegraphics[width=0.48\textwidth]{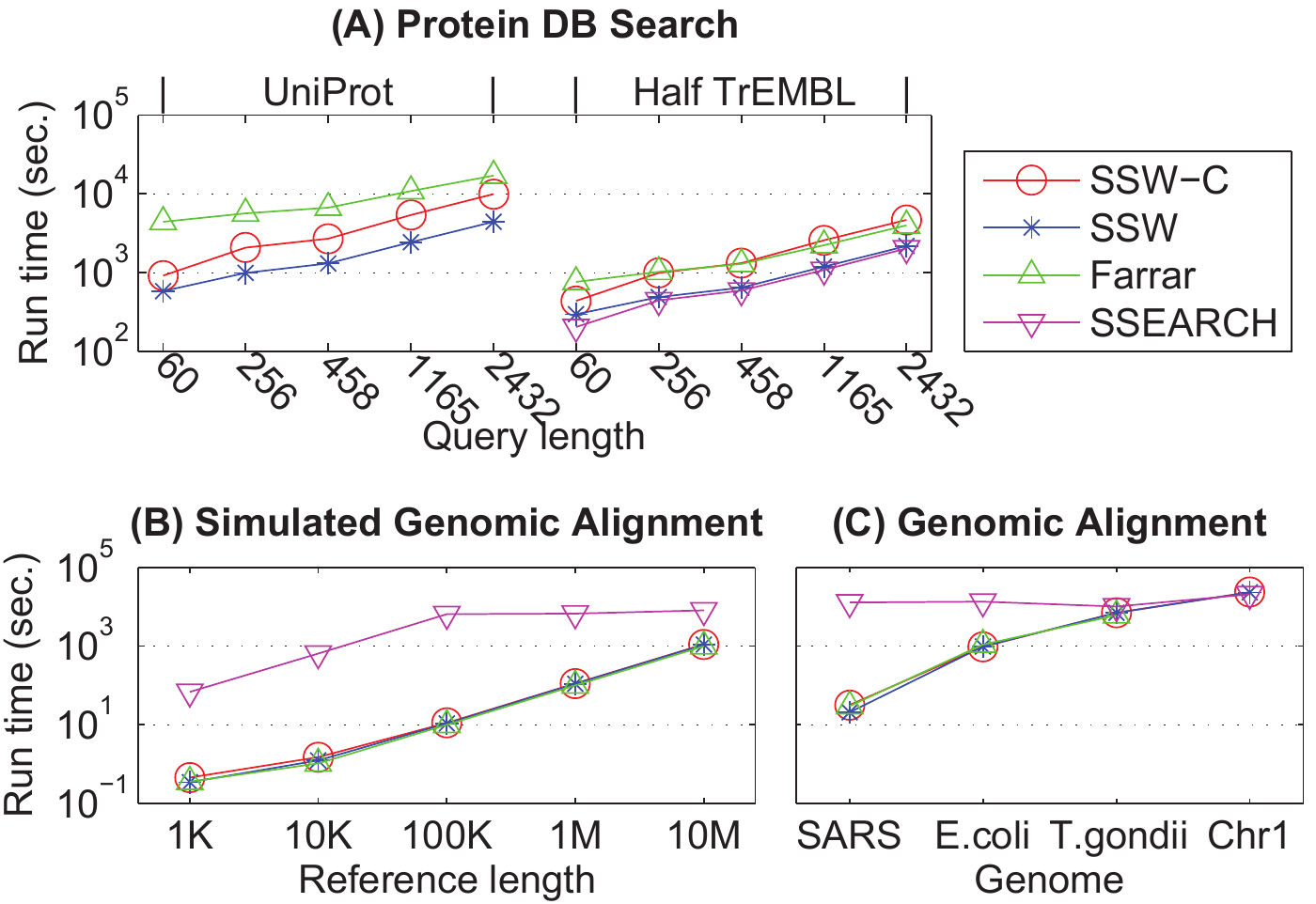}
\caption{Running time for different SW implementations. Log-scaled running time is shown on the y-axis for SSW without (blue) and with (red) detailed alignment, Farrar's implementation (green) and SSEARCH (pink). (A) Running times are shown for searching protein sequences against the full UniProt database (left) and one half of the TrEMBL database (right). All SW implementations used the BLOSUM50 scoring matrix with gap open penalty -12 and extension penalty -2. (B) The running time of aligning 1,000 simulated Illumina reads to human reference sequences of various lengths; and (C) of aligning 1,000 real sequencing reads to various microorganism genomes and the human chromosome 1 are shown. Farrar's implementation cannot handle long sequences as human chromosome 1, so its corresponding running time is not shown here.}\label{fig2}
\end{figure}

We compared SSW's performance (with and without returning the detailed alignment, SSW-C and SSW, resp.) to Farrar's accelerated SW and SSEARCH (version 36.3.5c) on a Linux machine with 2GHz x86 64 AMD processors. We ran each program on a single thread. Since the optimal alignment scores for long DNA sequences given by SWPS3 are not consistent with others', we did not benchmark its running time here.\\
\indent To measure the speed of protein database searching, we aligned protein sequences of varying length (60-2,432 aa) against the UniProt Knowledgebase release 2012 01 (including Swiss-Prot and TrEMBL, a total of 6,751,887,709 aa residues in 20,662,136 sequences), for all four algorithms (see Figure \ref{fig2}(A)). Since SSEARCH did not return alignment results against the entire UniProt database, we were only able to test it against one half of the TrEMBL sequences (3,319,575,305 aa in 9,634,307 sequences). Our SSW algorithm was the fastest or equally fast to SSEARCH across the entire protein sequence length range we tested.\\
\indent To benchmark genome sequence alignment, we tested the programs with both simulated data and real sequencing reads. We selected 1Kb - 10Mb regions from human chromosome 8, and using an Illumina read simulator (http://www.seqan.de/projects/mason/) we generated a thousand 100 bp-long sequences from these regions. We then aligned these reads back to their corresponding reference sequences with each of the four algorithms and compared running time (see Figure \ref{fig2}(B)). For the comparisons on real sequencing datasets, we aligned four sets of a thousand reads representing three different sequencing technologies against four different reference genomes: (1) ABI capillary reads (1,388 bp average length) against the severe acute respiratory syndrome (SARS) virus (29,751 bp); (2) Ion Torrent reads (236 bp) against E. coli ($4.94\times10^3$ bp); (3) Illumina reads (100 bp) against T. gondii ($6.08\times10^7$ bp); and (4) Illumina reads against human chromosome 1 ($2.49\times10^8$ bp) as shown in Figure \ref{fig2}(C). The genome alignment performance of these programs under another set of SW parameters is shown in Supplemental Figure S1. These results indicate that even while returning a full optimal alignment and one suboptimal score, our SSW algorithm is just as fast as Farrar's accelerated version.
\section{Applications}
We are demonstrating the utility of our SSW Library as components within four different biologically meaningful applications. 
\subsection{Primary short read mapper}
To provide highly accurate alignments, most short-read mappers integrate a Smith-Waterman algorithm for a final ``polishing'' step. This step is especially important for aligning reads containing short insertions and deletions. Even though each SW run is short, it may be applied hundreds of millions of times within a single run of a mapper, and therefore even small inefficiencies result in wasteful resource usage. To quantify time savings with SSW, we compared the performance of our new method with the existing SW implementation within our own MOSAIK mapping program (http://bioinformatics.bc.edu/marthlab/Mosaik) that uses SW for the final read alignments. We found that the SSW library achieves a two-fold speedup compared with the current banded SW implementation within MOSAIK (see Table \ref{Tab:01}). Notably, MOSAIK is a multi-threaded program and thus SSW component in MOSAIK is running in parallel.
\begin{table}[!t]
\processtable{Table 1: Comparison of running time (seconds) between SSW and the original SW implementation within MOSAIK.\label{Tab:01}}
{\begin{tabular}{p{2.5cm}p{2.5cm}lll}\toprule
& Illumina 100 bp & 454 \\\midrule
banded SW & 70145.760 & 240535.730\\
SSW & 38927.380 & 98198.990\\\botrule
\end{tabular}}{We aligned three million Illumina 100 bp reads and one million 454 reads against the human genome.}
\end{table}
\subsection{Secondary short-read mapper}
Primary read mappers are often unable to map or properly align reads in structural variant (SV) regions, e.g. in regions of deletions, insertions, inversions, or translocations. Therefore, we developed a split-read aligner program, SCISSORS (https://github.com/ wanpinglee/scissors) to map reads across structural variation event boundaries (breakpoints), rescuing reads not mapped, or inaccurately mapped by primary mapping approaches. We used our SSW library to align ``orphaned'' or severely ``clipped'' fragment-end read mates (in the case of read pairs where one end-mate is aligned with high mapping quality, but the other mate is either unmapped or mapped with many unaligned or ``clipped-off'' bases) to the genomic regions indicated by the well-mapped mates' coordinates. Inclusion of the SW mapping routines available in our SSW library makes accurate and fast split-read alignment for SV detection possible. The split-read mapping functionality, using our SSW library, has also been implemented in our TANGRAM SV detection tool (https://github.com/jiantao/Tangram) used intensively in the 1000 Genomes Project to accurately detect MEIs (http://ftp.1000genomes.ebi.ac.uk/vol1/ftp/technical/working /20120815\_mei\_calls/BC/). In SCISSORS and TANGRAM, SSW is also called in multi-threaded way.
\subsection{Read-overlap graph generation}
To evaluate evidence for putative SV and large INDEL calls generated by assembly methods, we can employ a read-overlap graph generated by exhaustive pairwise alignment of a set of reads which co-localize in a specific genomic region. We tested the effect of the SSW library in this application against a standard SW implementation (https://github/ekg/smithwaterman). To do the speed comparison test, we generated read-overlap graphs for the genomic region 20:21026000..21027500 using 22543 reads (70 bp long) from 191 African samples in the 1000 Genomes Project dataset (See Supplementary information). The running time of the read-overlap graph generation program RZMBLR (https://github.com/ekg/rzmblr) embedded with an ordinary SW (1795.66 seconds) is about twice of that embedded with our SSW Library (973.02 seconds).
\section{Conclusion}
We developed and made available a fast SW library using SIMD acceleration. By returning not only optimal alignment score but also the actual alignment, as well as a secondary optimal or suboptimal alignment score, the SSW library is suitable for inclusion into other heuristic genomic sequence analysis programs requiring local SW alignment. The most significant utility of our development, however, is that our algorithms can be readily integrated into C/C++ software without modification  of the source code, accelerating development for larger software tools. SSW has already been adopted in four programs developed by our group: the primary read mapping tool MOSAIK, the split-read mapping program SCISSORS, the MEI detector TANGRAM, and the read-overlap graph generation program RZMBLR.
\section*{Acknowledgement}

\paragraph{Funding\textcolon} This work has been funded by grant R01 HG4719 from the National Human Genome Research Institute to GTM.

\bibliographystyle{natbib}
\bibliography{document}

\end{document}